\documentclass{article}

\usepackage[a4paper, margin=0.75in]{geometry}
\usepackage[utf8]{inputenc}
\usepackage[dvipsnames,table,xcdraw]{xcolor}
\usepackage[affil-it]{authblk}
\usepackage{appendix,verbatim,listings,braket,enumitem, kantlipsum,graphicx,multicol,mathtools,csquotes,amsmath,mathtools,url,subfiles,qcircuit,amsthm,amssymb,fdsymbol,rotating,caption,hyperref,fontawesome,xcolor,sectsty,arydshln,pdflscape,pdfpages}

\def\titlefont{\color{RoyalPurple}}
\sectionfont{\color{RoyalPurple}}
\subsectionfont{\color{RoyalPurple}}
\subsubsectionfont{\color{RoyalPurple}}

\let\OLDthebibliography\thebibliography
\renewcommand\thebibliography[1]{
  \OLDthebibliography{#1}
  \setlength{\parskip}{0pt}
}

\title{\titlefont \textbf{\huge Automated Quantum Software Engineering}
\\\LARGE - why? what? how? -}

\author{Aritra Sarkar}
\affil{Quantum Machine Learning research group, Quantum Computing division, QuTech,
\protect\\Department of Quantum \& Computer Engineering, Delft University of Technology, The Netherlands}

\date{}

\begin{document}

\maketitle

\begin{abstract}
    This article provides a personal perspective on research in Automated Quantum Software Engineering~(AQSE).
    It elucidates the motivation to research AQSE (why?), a precise description of such a framework (what?), and reflections on components that are required for implementing it (how?).
\end{abstract}

\let\thefootnote\relax\footnotetext{This article is based on a talk delivered at Vrije Universiteit Brussel, 1 December 2022.}

\section{AQSE: Why?}

Quantum computing~(QC) is increasingly gaining focus for stakeholders in high-performance computing.
A major research avenue is on maturing the quantum computing hardware, in terms of high-fidelity (decoherence, error rates of quantum operations) and scalability (number of qubits, connectivity).
While this has proved rather a challenging engineering feat, rapid strides were made in the last decade with a plethora of physical technologies capable of demonstrating controllable processing of quantum information.

With quantum devices making steady progress, the complementary field of quantum software engineering~(QSE)~\cite{zhao2020quantum,serrano2022quantum} is also gaining traction.
The field has its roots in the theoretical formulation of quantum information and the earliest quantum algorithms.
However, more recently, QSE has been rejuvenated in the light of being integrated within currently available quantum computing pipelines, and design methodologies from classical software to be compatible with near-future avatars of quantum processors that are envisioned in technological road-maps.
While the importance of the underlying hardware cannot be understated, it is important that these two fields of hardware and software progress in parallel to prevent a quantum winter scenario where we have large costly quantum devices with no clear understanding of what applications could benefit from it.

Currently, there are three approaches to quantum software development:
\begin{enumerate}[nolistsep,noitemsep]
    \item[$\mathcal{A}_1$:] Given the limited capabilities of a specific quantum computing hardware, what useful computing can be implemented on that system?
    \item[$\mathcal{A}_2$:] Given an industrial use case, how can it be solved using an existing quantum algorithm (with some possible minor tweaking thereof)?
    \item[$\mathcal{A}_3$:] Designing new quantum algorithms for novel scientific underpinnings (mostly for specific mathematical properties) inspired by the superior (or at least different) computing capabilities of quantum information.
\end{enumerate}
For $\mathcal{A}_1$, the focus is on extracting as much computation power as possible from noisy intermediate-scale quantum~(NISQ)~\cite{preskill2018quantum} devices.
The researchers advocate a hardware-software co-design~\cite{shi2020resource} approach for the current technology readiness level~(TRL) of QC.
This involves diluting the abstraction layers of the quantum accelerator stack.
While this would help to justify the research funding in quantum computing by demonstrating state-of-the-art proofs-of-concept implementations, these highly tuned pipelines become difficult to scale and design.
In $\mathcal{A}_2$, researchers advocate adhering to strict abstraction layers~\cite{bertels2020quantum}, with separation~\cite{bertels2021quantum} of concerns between the challenges of hardware~\cite{leymann2020bitter} and software.
This principled fashion of organizing the research produces modular designs that are hardware-aware-yet-agnostic and are better aligned with the aims of QSE.
While in $\mathcal{A}_3$, the involvement of implementation, either on proof-of-concept QC simulators or real quantum processors, are minimal.
The focus is on specific mathematical properties, their proofs of correctness, and derivations of resource complexity bounds.

To make this distinction clear, examples of typical problems addressed by these approaches would be: (1) solving a pixelated case of handwriting recognition on a specific QPU with a specific hardware gate set, (2) implementing a pipeline for satellite image processing using one quantum convolution layer on a neural network architecture, (3) proving if quantum computing would provide a speedup for the 4-color theorem.
It can be easily appreciated that there exists a considerable gap between these approaches - both in their aim and the level of expertise required to address them.

At this juncture, in this article, we explore promising research directions that will aid in the advancement of QSE.
More specifically, we address some of the major problems that the field of quantum software engineering faces:
\begin{itemize}[nolistsep,noitemsep]
    \item Quantum mechanics is counter-intuitive to human cognition
    \item The barrier to entry for quantum algorithm development requires very different training than classical software developers
    \item Coverage of statistical testing is not scalable due to exponential state space, and inspecting intermediate states is not feasible due to no-cloning 
    \item It is not possible to deploy realistic problem use cases on either quantum processors or quantum computing simulators
    \item Similar to data-driven deep learning, the hybrid-quantum-classical algorithms based on variational principles are not interpretable
\end{itemize}
In general, there is a need to reduce the barrier to entry for assessing the impact of QC for a use case.
This can be bridged either with training or by automation.
Various educational and industrial institutions are now investing in training the next generation of the quantum workforce~\cite{aiello2021achieving} via courses, workshops, hackathons, tutorials, popular science articles, etc.
The latter is a rather interesting research venture and will be the focus of this article.

A primary motivation towards automation is the counter-intuitive nature of the semantic understanding of a quantum algorithm.
In typical graduate-level courses, the formalism of quantum mechanics and quantum information is introduced.
These form the basis for advanced courses and research in quantum algorithms.
However, it becomes clear that a phenomenological perspective of quantum algorithms is not possible in the same sense as courses like computer architecture and organization, Boolean logic, or digital logic design courses are internalized.
While superposition can be understood as multiple parallel threads of execution, and projective measurement can be understood as a weighted random selection of the basis states, (similar to how these are implemented in QC simulators), this is not enough.
Gaining quantum advantage from algorithms depends crucially on orchestrating interference between those threads such that the non-solutions destructively interfere and thereby increase the amplitude of the solution states.
Such insights often depend on serendipitous moments for skilled researchers~\cite{shor2022early}.

Does that imply, understanding the benefits of QC and building QC-based software solutions would remain the forte of a small circle of researchers?
This is the core motivation behind AQSE.
In what follows, we will define what is Automated QSE and contrast it with similar approaches.
Thereafter, we will list and reason about some promising building blocks that will most likely be required to construct such a framework.
\section{AQSE: What?}

Let us define automated quantum software engineering~(AQSE) as: `a framework capable of synthesizing a quantum computing solution for a given application.'
The deliberate vagueness will be discussed and gradually refined in this section.
At its finest form, AQSE would take in user requirements and produce a quantum computing implementation that would be a valid solution that the user can plug into an existing software pipeline and reap the benefits.
With that moonshot in mind, let us understand two important aspects of AQSE.

\subsection{Usability of the framework}

Based on our motivation behind AQSE, the AQSE framework must conform to ease of use.
We will consider two aspects of ease: the user interface and the level of vagueness/rigor in the problem specification.

The barrier to entry to the use of software can be frugally reduced by having a graphical user interface~(GUI).
The evolution of most software bear testimony to this trend, from operating systems to programming environments.
While most application software has GUI, visual programming languages~(VPL) have not been as popular.
Tools exist to easily design such interfaces for a code (e.g. in Python) at the back end.
Current quantum tools are mostly developed by researchers for fellow researchers with considerable backgrounds in setting up programming platforms.
Thus, efforts on these are often considered superfluous.
An intuitive user interface would go a long way in lowering the barrier to entry.
A few commercial/educational quantum platform providers are considering this more seriously.
These include qBraid, Strangeworks, IBM Quantum Composer, Quantum Inspire, Elyah, Notate~\cite{arawjo2022notational}, Quirk etc.
However, there is a crucial difference between these and the AQSE requirements.
These platforms aim to perfect a quantum integrated development environment~(IDE), aiding researchers in setting up a cloud computing environment, interfacing with various quantum hardware and simulator platforms, visualizing the results, and managing the execution logs.
We propose focusing on a Low Code, and eventually a No Code Development Platform~(NCDP) for AQSE.

The problem specification interfaces the intent of the AQSE user with the AQSE engine.
NCDP alone would not make quantum accelerators more accessible if it involves drag-and-drop unitary gates, as with all current QC VPLs.
Thus, this involves a different modality of AQSE.
The problem specification should abstract the details of quantum information processing and focus on the functional or behavioral problem definition.
Since the quantum details are no longer visible to the user, the interface should not look very different from similar tools on classical computing platforms.
Thus, in many aspects, it will be similar to a no-code AI or AutoML.
These tools decouple programming languages and syntax from logic and instead take a visual approach to software development to enable rapid delivery. 
No-code AI with the additional capability of reasoning in quantum logic and synthesizing quantum software is the vision of AQSE.

Do such tools exist?
Certainly not in the quantum software engineering space.
NCDP is more common for simple situations like web development, mobile apps, and game logic (visual scripting).
An intermediate solution would be graphical node/flow-based programming interfaces like Simulink.
The blocks can be specified at various levels of abstraction, e.g., a database query application, a quantum search algorithm, a Grover diffusion block, a multi-controlled Z gate, or a native gate/pulse for specific quantum hardware.
We will delve more into these levels in the next section.

\subsection{Assessment of applicability}

The intentional software~\cite{simonyi2006intentional} development paradigm, for better or worse, abstracts away the quantization of the desired solution.
Thus, it is paramount to understand when quantum computation is useful in the first place, based on the user specification.
In the broadest sense, this in itself is the core business idea of many consultancy companies in the quantum technology space.
Of course, AQSE will not be able to be so versatile, and such a feature will only apply to a well-specified problem definition.

There is some well-understood domain knowledge that can aid in this process.
Quantum computation is among the only known violation~\cite{bernstein1993quantum} of the complexity-theoretical Church-Turing thesis~(CTT) that is allowed by our current laws of physics.
There exist the complexity class, called bounded-error quantum polynomial time (BQP), that includes problems that are faster on a quantum model of computation (typically proved using a quantum Turing machine)
However, there are a few subtleties that need to be unpacked in such theoretical underpinning:
\begin{enumerate}[nolistsep,noitemsep]
    \item The corresponding classical complexity classes are bounded-error probabilistic polynomial time (BPP) and the polynomial time (P) classes that are efficient on a classical probabilistic/deterministic Turing machine, respectively. Thus, the focus of studying BQP problems is to rather identify problems in BQP$\backslash$BPP or BQP$\backslash$P region of computational time complexity. Our knowledge of such problems includes only a few examples, although they are the shining gems of quantum algorithms. Some of the early quantum algorithms like Deutsch-Josca, Bernstein-Vazirani, Simon's problems, Forrelation are about mathematical properties. Algorithms with more practical motivations include Shor's discrete logarithm, Shor's factorization, and the HHL algorithm for solving linear equations.
    \item Quantum computation does not solve an expanded set of functions, i.e., they are at the same degree of Turing computability. This means it is not a strict violation of the CTT, only of its extended version. This allows any quantum computation to also be expressed at classical computation, which forms the basis of QC simulators. 
    \item There are many classical universal models of computation, e.g., Turing machines, cellular automata, Post machines, lambda calculus, Wang tiles, etc. These models are equivalent to each other within a polynomial time overhead. Similarly, there are universal models of quantum computation, like quantum Turing machines, quantum cellular automata, quantum lambda calculus, adiabatic quantum computing, measurement-based quantum computing, and the canonical quantum circuit/gate model.
    These are all related in similar ways to each other.
    \item Points (2.) and (3.) mean any requirement specified to AQSE can be translated to both a classical and quantum implementation. The code structure at the computability level cannot guide the choice of a quantum implementation, which makes such a choice difficult. However, it also makes it interesting, as not each code needs to be assessed more intelligently in a broader context to understand its suitability of quantum acceleration. For example, an arithmetic operation would not provide a speedup by translation to reversible logic but becomes imperative if it is part of a quantum algorithm that manipulates superposition states.
    \item An awful lot of industrial and socially relevant computational issues can be formulated as problems that belong to the non-deterministic polynomial time (NP) class (or rather strictly in NP$\backslash$P class). It is known that QC will not be able to solve NP problems efficiently (i.e., in polynomial time) under realistic assumptions (e.g., P$\ne$NP). 
    \item The quantum Grover search provably provides a quadratic speedup for unstructured database search. Almost any problem can be posed as a search problem over a solution domain, e.g., factoring can be a search over numbers that, on multiplication, equals the result. Similarly, all problems in the NP class can be posed as a search problem based on the constraints' satisfiability~(SAT) since SAT is NP-complete. 
    \item Points (5.) and (6.) are the main reasons why we witness such proliferating attempts to formulate NP-hard problems as quantum algorithms. While these do not aim for an exponential speedup of Point (1.), just solving on a quantum model might allow speedup because it is a different form of computational automata. This latter case is particularly the motivation for quantum annealing (where quantum tunneling can be beneficial for some specific optimization landscapes over thermal fluctuations) and Boson sampling. Thus, understanding the formal model of computation is important, as a quantum search on adiabatic quantum computing~(AQC) would perform badly compared to a quantum circuit model, similar to how simulating Game of Life on a Turing machine would perform poorly compared to a cellular automata substrate.
    \item Besides these complexity theoretic viewpoints, it is important to realize that there are many problems where time complexity is not the major driver. Such problems are particularly studied in machine learning~(ML) and focus on space complexity~\cite{ventura1998quantum}, generalization, representation capacity, pattern recognition~\cite{schuld2019quantum}, etc. Thus, holistically the quantum solution needs to be assessed against other computational resources and metrics like memory requirements, convergence speed, solution accuracy, etc.
\end{enumerate}

These considerations imply that, though the interface of the No-code AI of AQSE would look welcoming to users, the underlying automation engine needs to be founded in rigorous mathematical principles to even assess the applicability, let alone design the quantum solutions.
At a superficial level, this seems as if they are at odds with each other since the functional level description is about abstracting away resource details, while resource estimates are crucial to assess the applicability.
This is the core innovation that AQSE addresses via increasing rigorous levels of abstraction and specification.
It will become clear in the following section that understanding the resource advantage of quantum software and synthesizing quantum software from requirements are the same thing from two different perspectives.
\section{AQSE: How?}

Having presented the overall goal of AQSE for the external interface and internal engine, in this section, we will delve deeper into components that will be necessary for the internal engine.

This is perhaps the right moment to clarify that AQSE is not an esoteric and novel venture.
There have been some attempts in the past to automate quantum algorithm design.
As early as 2004, a book~\cite{spector2004automatic} titled Automatic Quantum Computer Programming discusses evolutionary approaches for discovering novel quantum algorithms.
More recently, in the ongoing quantum computer engineering revolution, a few academic and commercial groups are pursuing this same goal.
Three of the most notable groups are discussed here, however, there are many individual researchers whose theses are aligned with AQSE.
Munich Quantum Toolkit~(MQT) from an academic group at the Technical University of Munich includes a set of tools relevant to design automation for QC~\cite{zulehner2020introducing}.
Most relevant to AQSE is the MQTProblemSolver~\cite{quetschlich2022towards} and SyReC Synthesizer~\cite{adarsh2022syrec}.
Horizon Quantum Computing is a company founded in 2018 in Singapore that aims to democratize quantum computing applications for businesses by removing the need for quantum algorithms knowledge for software developers. 
It features a compiler that automatically constructs quantum algorithms from classical code.
Their patent~\cite{fitzsimons2021systems} and public presentations reveal a layered approach for various levels of synthesis.
Another company, Classiq, based in Israel, aims to revolutionize the process of developing quantum computing software.
Their software platform transforms high-level functional models into optimized quantum circuits.
This allows quick development of large qubit circuits and execution on any gate-based system.
They hold a couple of patents~\cite{classiq1,classiq2,classiq3} on their offering that concerns AQSE.
The core of their inspiration, like MQT, is to repurpose methods from classical CAD in VLSI logic design for quantum circuits.
Discussions on the specifics of these tools and others (like AlgebraicJulia~\cite{brown2022compositional}, DisCoCirc~\cite{coecke2021mathematics}, SilQ~\cite{bichsel2020silq}, AdaQuantum~\cite{nichols2019designing}, Wolfram Quantum Framework~\cite{gorard2021zx}, etc.) will be introduced in the respective components.

\subsection{Refinement levels}

The AQSE engine is essentially a stack of abstraction layers connecting an implementation to a user intent.
Here we present some components that will be crucial to develop AQSE.

\subsubsection{User intent to application specification}

Foremost, AQSE requires inputting the user intent.
Very broadly, this can be classified as (i) an objective or (ii) creative/novelty.
The latter case involves the automated discovery of quantum algorithms and their corresponding purpose.
This is a rather niche field and has mostly been explored in the context of robotics.
However, a similar framework can be applied to (quantum) program synthesis.
We will not discuss this here in detail and will focus on objective-driven AQSE.

The specification language for the objective determines the level of vagueness allowed.
High vagueness translates to larger solution space.
However, it also gives a certain degree of freedom, and any solution from the larger space is assumed to satisfy the requirement of the user.
In many synthesis frameworks, the specification is iteratively refined by presenting behavioral examples to the user, eventually scoping the correct bounds of the problem space.

The key aspect of interfacing with the user, as discussed in the previous section, is a classical NCDP that abstracts quantum information processing as well as programming syntax.

\subsubsection{Formalizing application specification}

At the high end of the vagueness spectrum, we already witness the proliferation of natural language-driven coding, e.g., using OpenAI's Codex~\cite{chen2021evaluating} based on a modification of GPT framework.
These are based on an enormous corpus of training data, which might not be readily available for quantum computation.
However, Codex and Qiskit have already shown some initial promising results.
A more sustainable and explainable abstraction would be to refine the natural language to a formal specification language which can be further processed downstream in a controlled fashion.

A slightly higher structure is obtained in specification based on pseudo-code or LaTeX.
LaTeX to Python code converters already exists for mathematical equations.
Such tools can be handy specifically for optimization use cases based on SAT/SMT solvers, which can readily be translated to QUBO and thereafter to variational algorithms like QAOA~\cite{bako2022near} or quantum annealing.
These can be integrated with frameworks like SilQ to enhance user accessibility.
Similarly, software design frameworks like UML have also been extended to Q-UML~\cite{perez2021modelling}, which can be integrated into AQSE's NCDP.

AQSE should also retain the current level of specification at the QASM or embedded domain-specific language~(DSL) level.
These include cQASM, OpenQASM, Qiskit, Q\#, OpenQL, etc.
In today's NISQ era, this would also allow more trained users to specify non-functional requirements, like noise level, connectivity topology of hardware, qubit multiplicity, etc.
However, the focus on AQSE is on future generation of quantum processors where the end-user is not concerned with these low-level details, and can focus on algorithm development.

The right level of requirement specification is, of course, formal specification languages, like Z notation (Object Z, Z++)~\cite{cartiere2022formal} or B method.
Another alternative that has a low level of obscure syntax is logic programming languages like Prolog.
However, these tools have a steep learning curve and are unknown to most software developers.
Thus, the refinement to this level of abstraction must be encapsulated by AQSE.
We need to derive two things at this level, the functionality, and how to test/qualify it and thus bind the intention and validation aspects.
The validation can either be analytical or a set of test examples.
The AQSE NCDP would output a classical formal specification of the user requirements.

\subsubsection{Formal specification to formal logic}

Formal specification languages can be easily refined to 1st order predicate logic, or proof obligations for interactive theorem provers~(ITP) like Coq, Aqda, LEAN, etc.
The crucial aspect at this stage is to choose the formal logic to express the axioms, theorems, and the validity of proof entailments.
While classical or intuitionist logics are typically the default choices, in quantum it is worthwhile to use linear logic~\cite{girard1987linear}, which nicely captures the no-cloning of quantum information.
The corresponding language to express the logic is the dagger-lambda calculus~\cite{atzemoglou2014dagger}.
However, this needs further exploration and consideration for other candidates like modal logic, temporal logic, CTL* and many-value logic (e.g., paraconsistent logic~\cite{goertzel2021paraconsistent}).

\subsubsection{Solution representation}

The synthesized artifact that is gradually constructed by the AQSE engine needs to be represented and stored.
Typically in formal logic, proofs are represented in normal form (natural deduction) or tree form.
Based on the logic used, other options like Kripke semantics, sequent form, etc. can also be explored.
A natural way to store and explore proofs is via proof nets~\cite{girard1989proofs}, in a graph data structure.
This allows easy manipulation, rewriting, probabilistic reasoning, etc. using already well-developed libraries in most programming languages.
Proof trees can alternatively be replaced by abstract syntax trees~(AST) or abstract semantic graphs~(ASG).

\subsubsection{Search space representation}

Only a potential solution is represented as a proof nets (or, AST/ASG), it can be related to other solutions.
This can be via a meta-graph structure, where the edges represent the relation between the solution (e.g., one requires a qubit less, while the other requires 5 CNOT gates more).
In expressing relations (instead of functions), it is often desirable to represent a group of solutions that has a certain property.
Thus, we suggest using a generalized meta graph with hyper-edges, as the search space representation.
Similar constructs are used in the Wolfram Quantum Framework~\cite{gorard2020zx} and the OpenCog Hyperon~\cite{goertzel2021reflective} AGI cognitive architecture.

\subsubsection{Synthesis method}

The synthesis of valid solutions and their corresponding estimation of computational resources is the core of the AQSE engine.
There are various methods for program synthesis (or in this case, proof/AST/ASG synthesis).
Some of the most promising for our purpose are listed here, in increasing order of sophistication required to implement them:
\begin{itemize}[nolistsep,noitemsep]
    \item Exhaustive enumeration: easiest to implement, however, the entailment graph grow exponentially and becomes intractable beyond small instances. Formalisms like Nielsen geometry~\cite{nielsen2006quantum} and uncomplexity metric~\cite{brown2021effective} need to be incorporated to guide the search process.
    \item Pruned search:
    \begin{itemize}[nolistsep,noitemsep]
        \item Template-based meta-programming: can be used for small instances (or, holes in program synthesis) to fine-tune a code this is already very close to an acceptable solution. However, this is not specifically the goal of AQSE.
        \item Evolutionary approach: genetic programming~\cite{spector2004automatic} based solutions can be easily integrated with ProofNet/AST/ASG using linear logic. Other evolutionary approaches like novelty search~\cite{lehman2011novelty} and gene expression programming might also be useful.
        \item Neural networks: artificial neural networks~(ANN) and deep learning has been successfully applied in many cases on program learning, including the recent success of AlphaTensor~\cite{fawzi2022discovering}.
        \item Neuro-evolution: algorithms like NEAT~\cite{stanley2002evolving} and its later upgrades decouples the hyperparameter tuning and neural architecture search to an evolutionary heuristic.
        \item Neuro-symbolic approach: trades off between the explanability of symbolic AI with the efficiency of ANNs, and are specifically suited for symbolic regression tasks like theorem proving.
    \end{itemize}
    \item Automated theorem proving~(ATP): these proofs are basically based on the Curry-Howard correspondence~\cite{baez2010physics} between mathematical proofs and programs on universal automata.
    \begin{itemize}[nolistsep,noitemsep]
        \item Deductive proofs: are typically what is common in ATP. Some corresponding quantum solutions for expressing quantum proofs already exist, like QWIRE~\cite{paykin2017qwire}, SQIR~\cite{hietala2020proving}, CoqQ~\cite{zhou2022coqq}, LQP, QHL, etc.
        \item Categorical quantum mechanics: is a diagrammatic language for formal reasoning in quantum information. Tools like DisCoCirc, ZX-calculus, Quantomatic, and Catlab.jl can be used for the refinement of ASG to quantum programs. Research is needed in computational category theory for applied sciences (in contrast to applied category theory, which focuses on formalizing and understanding applied sciences rather than proactive computational development).
        \item Probabilistic proofs: allows uncertainties~\cite{nori2015efficient} in the user specification to trickle down to formal synthesis in a controlled manner. Tools like Markov logic networks~\cite{richardson2006markov} and probabilistic logic programming~\cite{goertzel2008probabilistic} (e.g., ProbFOIL~\cite{de2015inducing}) can be upgraded to incorporate quantum logic.
        \item Inductive proofs: allow generating solutions from incomplete specifications. Similar concepts have been studied in the quantization~\cite{arunachalam2017guest} of probably-approximate correct~(PAC) in learning theory. However, inductive tactics and approximations~\cite{bornholt2015approximate, andriushchenko2021paynt} need to be incorporated in quantum formal proofs.
    \end{itemize}
    \item Reinforcement learning: allows learning the solution given access to the environment. For AQSE, the environment can be a real quantum device, a quantum computing simulator, or the set a corpus of input-output training sets (called, programming by example). Similar techniques are explored in Hamiltonian learning, projective simulation~\cite{saggio2021experimental}, quantum knowledge seeking agent~\cite{sarkar2022applications}, and quantum photonics setups like MELVIN and AdaQuantum.
\end{itemize}

We expect that a future implementation of an AQSE framework would most likely be a subset of these features.
However, it is crucial to comprehensively evaluate~\cite{gulwani2017program} the applicability of at least (and most likely, more of) these techniques in the context of AQSE.

\subsubsection{Formal verification}

Formal verification is baked in the AQSE engine and represents a complementary research direction to stochastic verification~\cite{wang2021quito}.
It is, however, important that the formal proof of correctness also remain inspectable and interpretable to end users.
Tools that translate proofs to natural language, e.g. Coqatoo~\cite{bedford2017coqatoo}, can be used for such purposes.

\subsubsection{Hardware specific non-functional requirements}

Most available quantum processors are universal, in the sense that they have a defined set of native quantum universal gates.
However, the exact implementation cost depends on various factors like decoherence time, gate errors, qubit connectivity topology, control system multiplexing, etc.
In this article, we focused on the functional aspects of AQSE, with a theoretical pareto-optimization of quantum computing resources.
Low-level cost estimation are available in many available compilers which can be plugged into AQSE's synthesis cost estimator to specialize the framework for target hardware.



\newpage

\section*{... the way ahead}

\begin{center}
``We can only see a short distance ahead, but we can see plenty there that needs to be done."
\\(Alan Turing)
\end{center}

This survey of the current state of quantum software engineering and the need for automation is intended to not only scope out the research field around the author's interest in automated discovery of quantum algorithms, but also a pragmatic research proposal and call-to-action for multidisciplinary researchers working on allied fields.
AQSE is by no means an easy task. 
While AQSE can be critiqued as being futuristic and not applicable to current NISQ devices, it is important to look beyond the immediate needs and extrapolate the growth and needs of the quantum software industry a decade from now.
More specifically, AQSE would lead to the exploration of the limits of intelligence systems in contrast to what humans can achieve.

\section*{Acknowledgments}

I would like to thank Kobe Wullaert and Benedikt Ahrens for the discussions on category theory; Radouane Oudrhiri for introducing me to linear logic and formal methods in software engineering;  and Medina Bandic and Sebastian Feld for sharing the vision of machine learning in quantum computing development.

\bibliographystyle{unsrt}
\bibliography{ref.bib}





\end{document}